\documentclass[]{spie}  
\usepackage[]{graphicx}

\title{Rapid alerts for following up gravitational wave\\ event candidates} 

\author{Peter S. Shawhan\supit{a}, for the LIGO Scientific Collaboration and Virgo Collaboration
\skiplinehalf
\supit{a}The University of Maryland, Physics Dept.\ and Joint Space-Science Institute,\\
College Park, Maryland, USA
}

\authorinfo{Author E-mail: pshawhan@umd.edu}

\begin{document} 

\begin{center}
{\bf Cover page with copyright notice required by SPIE }
\end{center}

\vspace{0.75in}

{\flushleft
{\it This document is:}

Peter S. Shawhan, for the LIGO Scientific Collaboration and Virgo Collaboration, ``Rapid alerts for following up gravitational wave event candidates'', Observatory Operations: Strategies, Processes, and Systems IV, Alison B. Peck, Robert L. Seaman, Fernando Comeron, Editors, Proc.\ SPIE 8448, 844825 (2012).

Copyright 2012 Society of Photo-Optical Instrumentation Engineers. One print or electronic copy may be made for personal use only. Systematic electronic or print reproduction and distribution, duplication of any material in this paper for a fee or for commercial purposes, or modification of the content of the paper are prohibited.

DOI: To be assigned
}

  \maketitle 

\begin{abstract}
Gravitational waves carry unique information about high-energy
astrophysical events such as the inspiral and merger of neutron stars
and black holes, core collapse in massive stars, and other
sources. Large gravitational wave (GW) detectors utilizing exquisitely
sensitive laser interferometry---namely, LIGO in the United States and
GEO 600 and Virgo in Europe---have been successfully operated in
recent years and are currently being upgraded to greatly improve their
sensitivities. Many signals are expected to be detected in the coming
decade. Simultaneous observing with the network of GW detectors
enables us to identify and localize event candidates on the sky with
modest precision, opening up the possibility of capturing optical
transients or other electromagnetic counterparts to confirm an event
and obtain complementary information about it. We developed and
implemented the first complete low-latency GW data analysis and alert
system in 2009--10 and used it to send alerts to several observing
partners; the system design and some lessons learned are briefly
described. We discuss several operational considerations and design
choices for improving this scientific capability for future
observations.
\end{abstract}


\keywords{Gravitational waves, LIGO, Virgo, GEO 600, multi-messenger, time-domain astronomy, afterglow}

\section{INTRODUCTION}
\label{sec:intro}  

The direct detection of gravitational waves has long been heralded as
a new observational tool for astronomy and astrophysics.  Indeed,
strong gravitational waves are surely emitted by binary systems of
neutron stars and/or black holes in close orbits, as already confirmed
indirectly by radio observations of pulsars in a few such systems in
our galaxy \cite{TaylorMeas,Weisberg2010,DoublePulsar}.  Ground-based
gravitational wave (GW) detectors can directly detect signals from the
final seconds to minutes of ``inspiral'' leading up to the merger of a
binary system
\cite{Peters,Last3min}.  Short-duration GW signals are also expected
to be emitted from the collapsing cores of massive stars
\cite{OttReview,FryerNew} and from perturbed neutron stars and black
holes \cite{KokkotasSchmidt}, although the dominant excitation and
emission mechanisms---and thus the strength of the waves---are much
less clear.  GW transients might even be produced by cosmic strings
\cite{DamourVilenkin2001} or other exotic sources.

Ground-based detectors are also seeking continuous signals from
non-axisymmetric spinning neutron stars
\cite{JKS,KnownPulsars,S5PowerFlux} and stochastic GW background
radiation \cite{MaggioreStoch,S5Stoch}, while pulsar timing campaigns
\cite{Parkes,EPTA,NANOGrav} are currently monitoring selected stable
pulsars to pick out the signatures of very-low-frequency GWs from
supermassive black hole binaries and other sources.  Someday,
hopefully, space-based interferometers similar to
LISA~\cite{LISAReview} will be launched to study the intermediate
frequency band and the galactic and extragalactic GW sources that
populate it~\cite{eLISAScience}.

While GWs still have not yet been directly detected, the detectors
operated up to this point have successfully collected data which have
been analyzed to set ever-improving upper limits on the rates and
strength of likely GW signals (such as Refs.~\citenum{S6VSR23CBC}
and \citenum{S6VSR23Burst}).  Major upgrades currently underway
should finally enable the detection of GW events later this decade,
and the GW science community is now preparing to take full advantage
of the new capabilities.  This presents an excellent opportunity for
GW scientists and astronomers to collaborate to obtain multi-messenger
observations of transient events through their electromagnetic (EM) as
well as GW emissions, regardless of which messenger first detects a
given event.  The purpose of this article is to discuss the prospects
for EM follow-up observations triggered by the rapid identification of GW
event candidates.  First steps in that direction have already been
taken, and a broader future program is now being planned.

\section{THE GRAVITATIONAL WAVE DETECTOR NETWORK}
\label{sec:detectors}  

Direct detection of gravitational waves was first attempted using
resonant ``bar'' detectors \cite{WeberOrig,IGEC2003}---and the AURIGA
\cite{AURIGA} and NAUTILUS \cite{EXNAU} cryogenic bar detectors are
still operating---but the past two decades have seen the rise of
large-scale laser interferometers which achieve better sensitivity and
span a much wider frequency range.  The basic detection principle is
simple: a laser beam is split into two beams which travel along
perpendicular ``arms'' with mirrors at the ends, and interference when
the reflected beams are recombined indicates the minute differential
changes in the effective lengths of the arms produced by a passing
gravitational wave.  But the remarkably small amplitude of the GWs
reaching Earth, with a strain (fractional length change) of order
$10^{-21}$ or less, demands exquisite stability and control of the
laser and optical systems, mirror positioning and alignment, low-loss
optical components, careful management of sensor and actuator noise,
and nearly total isolation from ground vibrations and other
environmental disturbances.

Gravitational-wave ``observatories'' with kilometer-scale
interferometric detectors have been built and operated in the United
States (LIGO) and Europe (GEO 600 and Virgo), as described below.
Medium-scale prototypes have also been developed in Japan as a prelude
to building a large detector.  As these are effectively quadrupole
antennas, each detector has nonzero response to GWs arriving from
almost all directions, with direction- and polarization-dependent
``antenna factors'' between 0 and 1; and because GWs pass through the
Earth unimpeded, {\em all} operating detectors generally will respond to the
same incoming wave.  Allowing for direction-dependent arrival time
differences, the GW detectors can be used together as a coherent
network with multiple baselines.  Recognizing this fact, the
collaborations operating the different detectors have agreed to
combine and jointly analyze all of the data, and this cooperative
philosophy is expected to continue into the future.

As of this writing in mid-2012, the field is in the midst of a major
transition, as the first-generation detectors (briefly
summarized below) have ceased operating and new ``advanced'' detectors
are being constructed in their places.  We will again have a true
network of GW detectors operating later this decade, with roughly an
order of magnitude greater amplitude sensitivity than in the past.

\subsection{First-generation detectors}

The LIGO project \cite{LIGO1992,LIGORoPP} has built two observatories---on
the Hanford reservation in Washington and in Livingston Parish,
Louisiana---and has operated {\em three} detectors simultaneously: the
LIGO Hanford Observatory housed interferometers with 4-km and 2-km
arms, sharing a vacuum envelope, while LIGO Livingston has had a 4-km
interferometer matching the longer one at Hanford.  LIGO is primarily
funded by the U.S.\ National Science Foundation (NSF) through the
``LIGO Laboratory'' at Caltech and MIT, but the science mission of
LIGO is carried out by a much larger organization, the LIGO Scientific
Collaboration (LSC), which includes hundreds of scientists from the
U.S., Europe and elsewhere.  LIGO detector installation began in
1998 and a series of short ``science runs'' began in 2002,
albeit with limited sensitivity.  After further commissioning, the
LIGO detectors reached their design sensitivity goal in 2005 and
recorded data in the ``S5'' science run from 2005--7 \cite{LIGORoPP}.
With some enhancements (testing technologies to be used by the
advanced detectors described below), LIGO had an S6 science run
from 2009--10.

The GEO 600 detector \cite{GEO2008} (also known simply as GEO) has
been built by a collaboration of German and British scientists and is
sited near Hannover, Germany.  With folded 600-meter arms, GEO is
smaller than the other km-scale interferometers but has pioneered more
ambitious techniques for interferometry and mirror suspensions that
have informed the designs of later detectors.  The GEO 600 detector
and the scientists who work with it are fully part of the LSC (despite
the fact that only LIGO appears in the {\em name} of the
collaboration), and GEO 600 has often collected data in science runs
synchronously with LIGO.  Additionally, the GEO team generally tries
to operate the detector and record data whenever it is not being
actively upgraded or commissioned, in an ``AstroWatch'' mode to be
prepared for any remarkable event such as a supernova in our galaxy.

Virgo \cite{Virgo2012} is a 3-km detector located in Cascina near Pisa, Italy
with a design generally similar to LIGO's.  Begun as a French-Italian
project funded by the Centre National de la Recherche Scientifique
(CNRS) and the Istituto Nazionale di Fisica Nucleare (INFN) through
forming the European Gravitational Observatory (EGO) consortium, the
Virgo Collaboration now includes researchers from other European
countries as well.  The Virgo detector reached a scientifically
significant sensitivity level in 2007 and carried out its first
science run, VSR1, together with the last 5 months of the LIGO/GEO S5
run, enabling the first LIGO-Virgo joint
analyses~\cite{S5VSR1Burst,S5VSR1CBC,S5VSR1GRBBurst,S5VSR1GRBCBC}.
The next two science runs, VSR2 and VSR3, were synchronous with the S6
run; a VSR4 run followed after further commissioning of the new set of
mirrors installed between VSR2 and VSR3.

\subsection{Advanced gravitational wave detectors}

GEO is arguably the first ``advanced'' GW detector, with a number of
recent and planned technological improvements~\cite{GEOupgrades} in
laser power, signal recycling, and the use of ``squeezed light'' to
push below the normal quantum shot-noise level~\cite{GEOsqueezing}.
This program of upgrades is called ``GEO-HF'' since the main focus is
on high frequencies, above several hundred Hz.  GEO is currently
collecting data in AstroWatch mode with an uptime of about 70\%.
However, GEO's relatively high noise at low frequencies means that it
has rather limited sensitivity for binary inspirals and other
low-frequency sources.  A major step forward will come later this
decade with the completion of the Advanced LIGO and Advanced Virgo
upgrades, and the construction of the KAGRA detector in Japan.

Advanced LIGO \cite{AdvLIGO} is a nearly total replacement of the
instrumentation at the LIGO observatories with higher-power lasers,
larger mirrors, a dual recycling optical configuration with stable
recycling cavities, multi-stage passive and active mirror suspension
systems, and new control and readout electronics.  These changes are
designed to reduce the limiting noise contributions over the entire
frequency band by an order of magnitude compared to the initial LIGO
detectors, as well as extending downward to $\sim$$10$ Hz.  The use of
squeezed light is being studied as a possible enhancement.  Components
are currently being assembled and tested for three detectors, all 4 km
long.  The original plan was to install two detectors at Hanford and
one at Livingston, but there is now a fully developed proposal to
install one of the detectors at a new observatory to be built in
India, which will form additional long baselines and greatly improve
the ability of the network to localize sources in the sky.  The first
two detectors, at Livingston and Hanford, should be fully installed by
the end of 2013 and basically operational by 2014; following initial
commissioning, a first science run is likely in 2015.  After further
commissioning and tuning, the detectors should reach their design
sensitivity levels a few years later.  Assuming that the LIGO-India
plan is approved by the U.S.\ National Science Board and that Indian
funding for the new observatory is confirmed, the third detector will
be installed at the Indian site late this decade and should be ready
to start collecting science data around 2020.

The Advanced Virgo \cite{AdvVirgoDocs} design is similar to Advanced
LIGO and involves many of the same kinds of upgrades, including higher
laser power, larger mirrors, and dual recycling.  The existing Virgo
``superattenuator'' mirror suspensions \cite{superattenuator}, with
multiple passive stages, will continue to provide sufficient isolation
from seismic noise.  An Advanced Virgo Technical Design Report has
been completed, and detailed planning and scheduling are now underway.
The goal is to have a robustly operating detector in 2015 and to begin
collecting science data as soon as possible after that.

Funding for the Japanese KAGRA detector \cite{KAGRA}, formerly known
as the Large-scale Cryogenic Gravitational-wave Telescope (LCGT), was
approved in 2010.  It will have 3-km arms and a dual-recycled optical
configuration similar to Advanced LIGO and Virgo.  Uniquely, KAGRA
will be located underground, in the Kamioka mine in western Japan,
where seismic noise and gravity gradient noise are much lower than at
the surface.  It will also have sapphire mirrors cooled to 20~K to
reduce thermal noise, as has been demonstrated with the CLIO 100-meter
cryogenic prototype \cite{CLIO}.  An initial 3-km room-temperature
interferometer without recycling is expected to be operational by
2015, with the full cryogenic interferometer ready to start taking
data by 2018.

\begin{figure}
   \begin{center}
   \begin{tabular}{c}
   \includegraphics[height=5.5cm]{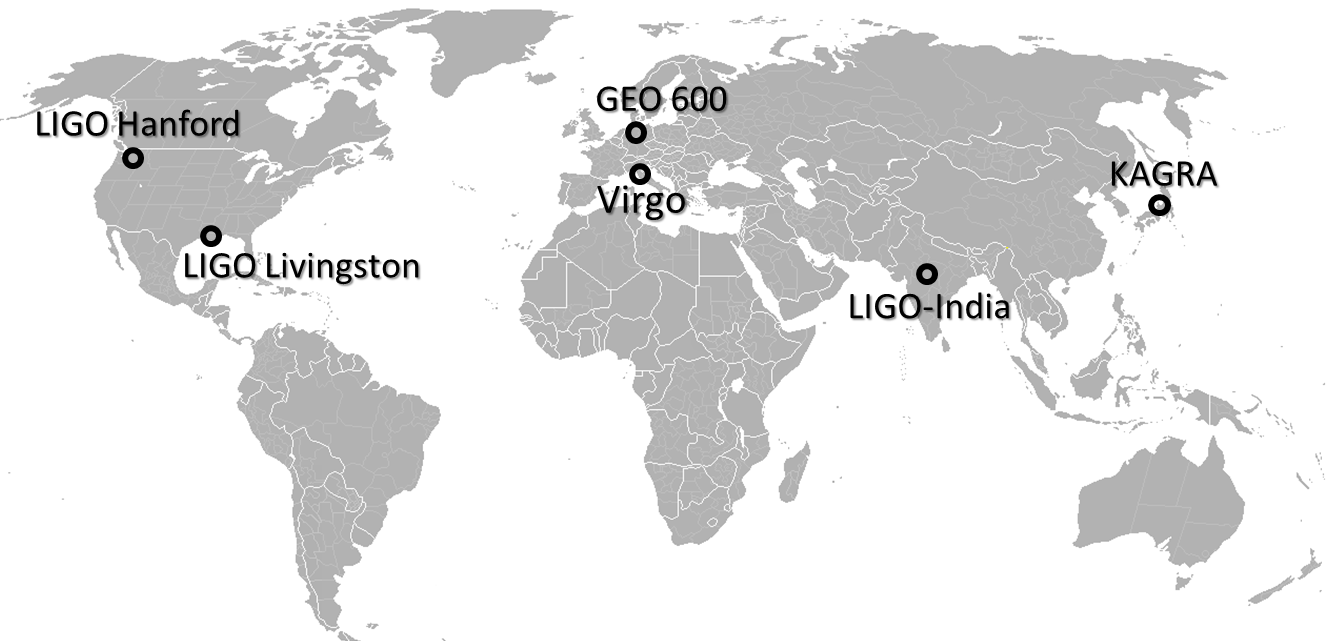}
   \end{tabular}
   \end{center}
   \caption[DetectorNetwork] 
   { \label{fig:IFOs} 
Locations of current and future kilometer-scale gravitational wave
detectors.  The GEO 600 detector is currently taking data; the Advanced
LIGO detectors at Hanford and Livingston and Advanced Virgo are
expected to begin taking data in 2015; KAGRA should operate in its
full optical configuration starting around 2018; and LIGO-India around 2020.
LIGO-India is contingent on final approvals and funding, and its exact
location has not yet been determined.}
\end{figure} 

Thus, within $\sim$3 years we should have two or three advanced
detectors collecting data with significantly better sensitivities than
the past network, with full design sensitivity and KAGRA following a
few years after that, and LIGO-India somewhat later; see
Fig.~\ref{fig:IFOs}.  The advanced detectors will be capable of
detecting signals in a volume of space $\sim$1000 times larger than
the initial detectors.  It has been estimated that the Advanced LIGO
and Virgo detectors will likely detect a few dozen binary merger
events per year \cite{RatesPaper}; although such estimates have large
uncertainties and the expectations for other event types are even less
well known, they motivate us to prepare now for a comprehensive
science program---and one that is not restricted to the GW signals
alone.

\section{CONNECTING WITH OTHER OBSERVATORIES}
\label{sec:connect}  

The most likely sources of detectable gravitational waves are all
highly energetic events; for example, the inspiral and merger of two
neutron stars releases $\simeq 5 \times 10^{53}$~erg of gravitational
binding energy.  While most of that energy is emitted in the form of
neutrinos and gravitational waves \cite{Piran1994}, a fraction of it
can go into electromagnetic radiation.  In fact, it
is believed that binary mergers are the progenitors of most short-hard
gamma-ray bursts (GRBs) (see, for example, Refs.~\citenum{Eichleretal}
and \citenum{Narayanetal}), with multiple observational features
supporting that hypothesis \cite{Bloom1999,Berger2010}.  Binary
mergers and other types of disruptive events may naturally power EM
emissions in the X-ray, optical, and radio bands through various
mechanisms.

Even the advanced GW detectors will be limited to the relatively
nearby universe by the detector noise floor; maximizing the detection
rate for GW events calls for setting event candidate selection
thresholds as low as possible while maintaining a very low
``background'' rate from noise fluctuations.  General searches using
GW data only must accept a candidate from any time or sky position.
{\em However}, having an observed EM transient counterpart with a
consistent time and sky position would greatly increase confidence in
a GW event candidate, and could thus be the key supporting evidence
that confirms the candidate, even if it was a bit too weak to be
cleanly distinguished from the background based on the GW data alone.
Furthermore, the EM counterpart may pin down the host galaxy and local
environment of the source, while comparison of the EM and GW
signals---e.g., relative timing and durations, relative energy
release, and polarization content---would provide unique information
about the identity and astrophysics of the progenitor.  For instance,
GW observations should eventually be able to confirm or rule out the
binary merger model for short-hard GRBs.

For several years, the LSC and Virgo have been carrying out deeper GW
searches around the times and sky positions of many astrophysical
``triggers'' (reported events) such as GRBs, magnetar flares, and
pulsar timing glitches; Refs.~\citenum{S6VSR23GRB},
\citenum{SixMagnetars} and \citenum{S5VelaGlitch} present some recent
examples.  Triggered GW searches are also in progress using
high-energy neutrino candidates and nearby supernovae as targets.
Such searches are typically about a factor of $1.5$ to $2$ more
sensitive than all-sky (un-triggered) GW searches, meaning that
sources can be detected up to twice as far away, extending the science
reach significantly.

Still, to gain the most from associating GW and EM events, we must
consider that not all EM transient events will be detected
promptly---if at all.  Excellent sky coverage is now available in
gamma rays and hard X-rays thanks to the Gamma-ray Burst Monitor (GBM)
\cite{GBM} on the Fermi spacecraft, which views all of the sky not
occulted by the Earth, together with the Interplanetary Network (IPN)
of gamma-ray instruments on several spacecraft \cite{IPN}, which see
the whole sky but with lesser sensitivity.  However, GRBs are
understood to be strongly beamed (see, for example,
Ref.~\citenum{FongBeaming}), so only a small fraction of their
progenitors can actually be observed via gamma-rays.  Systems which
launch gamma-rays somewhat off-axis to the line of sight, or which
produce only moderately relativistic jets (``failed GRBs''), may still
be detectable in the optical or radio bands as ``orphan afterglows''
\cite{Rhoads1997,Huangetal}.  Neutron star mergers are likely to also
produce fainter, isotropic ``kilonova'' light curves in the optical
band which are powered by the radioactive decay of elements produced
by $r$-process nucleosynthesis \cite{MetzgerKilonova}.  Because only a
small fraction of the sky is viewed at any given time by sensitive
optical and radio instruments, these transient signatures would only
be caught serendipitously---{\em unless} gravitational wave (or
high-energy neutrino) detectors can identify these events promptly and
accurately enough to tell telescopes where to point.

Data from the GW detector network can, in fact, be analyzed within
minutes to identify candidate events and reconstruct a sky map of the
likely position of each candidate.  This information can then be
passed to astronomers for follow-up imaging.  (The general strategy is
sometimes called ``LOOC-UP'' after an early pilot study
\cite{LoocUp}.)  We developed and tested such a system during the
2009--10 LIGO-Virgo joint science runs, and will support and improve
this capability for observing with the advanced GW detector network.
Below, we discuss the main characteristics of the past system as well
as some improvements envisioned for the future.

\section{OPERATIONAL CONSIDERATIONS FOR PROMPT FOLLOW-UP OBSERVATIONS}
\label{sec:operations}  

The overall goal is to identify transient events in the GW data
quickly, determine their sky positions as well as possible, and
communicate that information to observers with access to telescopes
for capturing images of the appropriate region(s) in the sky.  It is
desirable to minimize the latencies of all of those steps so that the
telescopes can catch a fading afterglow (if there is one) as early as
possible.  It should be noted, though, that other types of EM
emissions would take some time to appear, such as a kilonova light
curve which peaks after $\sim$1 day \cite{MetzgerKilonova}, or
synchrotron emission in the radio band \cite{NakarPiran} which would
spread over weeks to months.  Therefore, rapid alerts can support
both rapid and delayed follow-up observations.

\begin{figure}
   \begin{center}
   \begin{tabular}{c}
   \includegraphics[height=6cm]{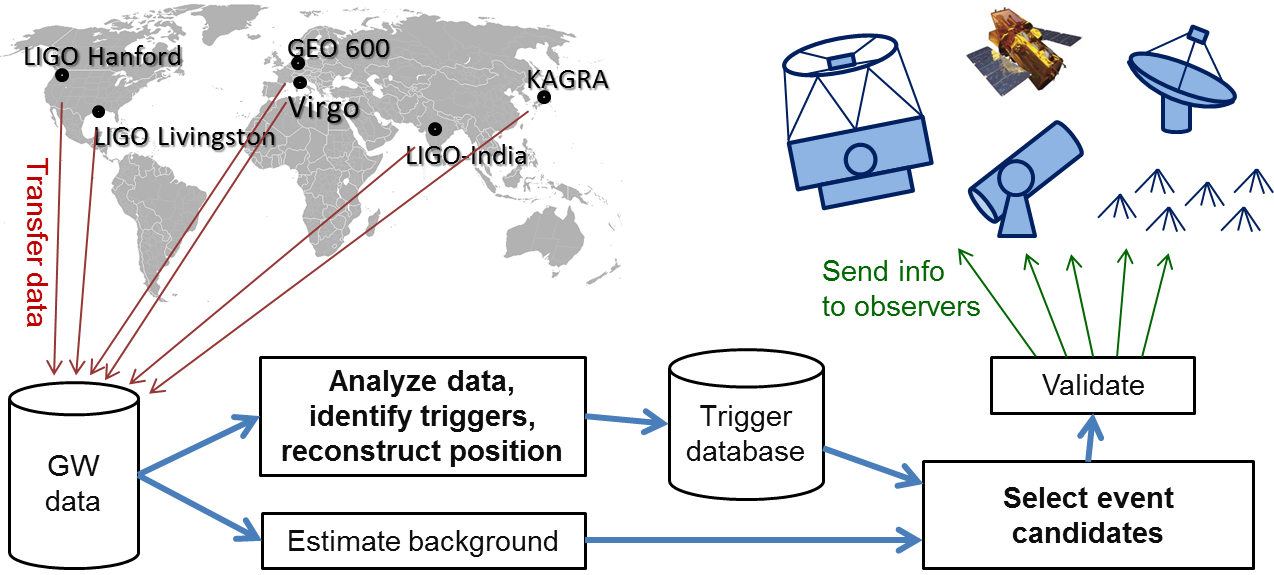}
   \end{tabular}
   \end{center}
   \caption[FlowChart] 
   { \label{fig:FlowChart}
Main steps in processing data from the GW detector network and rapidly generating alerts for follow-up observations.  ({\it Swift} image credit: NASA E/PO, Sonoma State University, Aurore Simonnet.)}
\end{figure}

For the 2009--10 LIGO-Virgo science run, we implemented a complete
mostly-automated data analysis, event selection and alert distribution
system, and passed alerts to several partner observers.  That system
and evaluations of its performance are described in
Refs.~\citenum{EMf} and \citenum{CBCLL}.  Figure~\ref{fig:FlowChart}
shows the main elements and general data flow through the system,
illustrated with the future network of advanced GW detectors.  In 
this section we discuss a number of operational considerations
based on our experience with the past system, along with some notes
about changes envisioned for the future.

\subsection{Data collection}  \label{subsec:data}

First of all, we need to have multiple GW detectors collecting data at
the same time with comparable sensitivity, because it is mainly the
difference in arrival times which tells us about the sky position of
the source.  (More about that in section \ref{subsec:position}.)  The
2009--10 follow-up program was active only when both 4-km LIGO
detectors and Virgo were all collecting science-mode data.  GW
detectors can operate day and night, but in practice they do not
operate all the time; bad weather conditions (causing elevated seismic
noise), occasional nearby human activities, instrumental problems and
scheduled maintenance periods all contribute to downtimes that have
historically been $\sim$10--30\%.  Thus, although science runs of the
various detectors are carried out synchronously when possible,
different subsets of the network will be collecting
science-mode data at any given time.

Data from all of the detectors needs to be calibrated quickly and
transferred to one or more computer centers to be analyzed together.
In 2009--10, calibration took $\sim$15--20 s and all of the GW strain
data was transferred to the main archive at Caltech and to the Virgo
observatory site in less than 1 minute, on average, ready to be
analyzed.  Data transfer latencies could perhaps be trimmed in the
future, although these have not been the limiting factor so far.

\subsection{Low-latency analysis}  \label{subsec:lowlatency}

Binary mergers are a natural target for rapid follow-up observations
because they are the best predicted GW source type and because there
are many models for EM emission at various wavelengths.  The
GW waveform from the inspiral phase leading up to merger mainly
depends on the masses of the two objects and can be calculated
accurately, at least for neutron stars and stellar-mass black holes
with relatively little spin.  This allows the data to be searched
using optimal matched filtering with a bank of templates.  In
2009--10, the data arriving at the Virgo Cascina site was promptly
processed using the MBTA algorithm to search for inspiral signals in
the three data streams \cite{CBCLL}.  Large output values from the
matched filters were recorded as GW ``triggers'' in a database, along
with reconstructed sky position information, within a few minutes.

To allow for other types of events which are not so well modeled (if
at all), we also apply GW ``burst'' search algorithms which are
designed to find transient GW signals in the data without making any
assumptions about the specific waveform.  The basic approach is to
test for signals above the baseline noise level with consistent
relative times, frequency and polarization content in the multiple
data streams.  In 2009--10, three different GW burst search algorithms
were used in parallel: coherent WaveBurst for arbitrary signals,
coherent WaveBurst with a linear polarization constraint, and Omega
Pipeline with a Bayesian coherent analysis stage\cite{EMf}.  These
algorithms responded best to different types of GW burst signals, and
each was set up to record outliers above a moderate threshold in the
trigger database, along with sky position information.  This
processing typically took about 5 minutes.

Background from noise fluctuations---in other words, the rate of
``false alarms''---is a critical consideration for any GW search
because very weak signals are being sought, and the statistical
properties of the noise are not known {\it a priori}.  The rate and
properties of background triggers can be estimated by time-shifting
data streams from the different detectors, which removes any
correlation from a real GW signal while sampling a different alignment
of random fluctuations and/or instrumental glitches (uncorrelated
since they are at widely separated sites).  Trying many different shifts and
re-analyzing an interval of data (or, for the MBTA coincidence
analysis, simply multiplying average single-detector trigger rates)
yields a background rate estimate with good statistical precision.

The next step is to select highly significant triggers from the
(un-shifted) analysis based on how unlikely they are to be random
background; this is quantified by an effective false alarm rate (FAR)
for the background distribution stronger than the trigger being
considered.  One difficulty is that the outlier {\em tail} of the
background distribution is crucial for evaluating the most interesting
triggers, but measuring that tail requires a considerable quantity of
data and/or many different time shifts.  With first-generation GW
detectors, at least, the background rate and properties change on
various time scales; thus there is an issue of deciding how best to
estimate the background for a given point in time.  In 2009--10, our
selection criteria involved multiple background estimates averaged
over different time intervals ranging from 10 minutes to several
hours \cite{EMf}.  That worked well enough that we were able to choose ``event
candidates'' to pass along to follow-up observers at approximately the
desired average rate, which was $\sim$1 observable candidate per week
for most partner telescopes and lower for certain others, including the
{\it Swift} XRT and UVOT\cite{EMfSwift}.

Finally, before sending an alert out to observers, we want to be sure
that it is not an obvious instrumental artifact to avoid spending
telescope time unnecessarily.  During the 2009--10 run, we used a
combination of automatic and manual checks on data quality and current
conditions at the detector sites at the time of each event candidate,
using the judgment of detector operators and scientists at the sites
together with an ``EM follow-up expert'' who was on shift at the
time.  About 1/3 of the event candidates were judged to be
questionable for some reason and were rejected.  This validation step
thus improved the ``purity'' of the rapid alerts to some extent, but
it added $\sim$10 to 30 minutes to the latency for sending out the
alert, dominating the total latency.  In the future, we may be able to
improve the automated checks to the point where we can remove the
manual validation stage and send alerts out more quickly.

\subsection{Position reconstruction}  \label{subsec:position}

Despite having baselines of thousands of kilometers, the GW network is
generally not able to reconstruct the sky position of a transient
candidate very well because the signals of interest are at low
frequency and low signal-to-noise ratio.  Position information comes
mainly from the difference in arrival times between detector sites,
and the resulting position uncertainty can be predicted approximately
in terms of the properties of the signal waveform
\cite{Fairhurst1,Fairhurst2}.  There is, however, additional
information in the relative signal amplitudes since the detectors are
oriented differently and therefore have different antenna factors for
both GW polarization components.  In any case, localizing events
well requires data from at least three detectors; with only two
detectors, we can still do a good search for GW events using
appropriate coincidence tests, but can only crudely localize
candidates along a ring on the sky.  Having four or more detectors in
the future will have the advantage that when one goes down, the others
can still localize sources well.

Achievable sky region areas have been studied analytically
\cite{Fairhurst1,Fairhurst2} and using Monte Carlo simulations
\cite{KlimenkoLoc,NissankeLoc}, and continues to be a topic of
interest.  It is difficult to summarize the results since they depend
strongly on the GW detector network operating at the time as well as
the waveform, amplitude and sky position of the GW signal.  With the
full future network of advanced GW detectors, sky error regions of a
few square degrees are predicted (e.g., for binary neutron star
inspirals), whereas earlier on with just the Advanced LIGO Hanford,
LIGO Livingston and Virgo detectors, error regions should typically be
tens of square degrees.  It should be
noted, though, that these estimates sometimes assume detectors with
equal sensitivities, which is unlikely to be the case.  Also, real
data analysis algorithms which have to deal with non-stationary noise
may not attain the ideal performance, and calibration uncertainties
can lead to systematic shifts in reconstructed positions.  Those
issues will need to be addressed carefully to ensure that telescopes
can be pointed accurately.

Simple time-of-flight triangulation with three detectors normally
picks out two areas of the sky; a fourth detector usually can resolve
the ambiguity.  The error regions can, however, be rather elongated
depending on their position in the sky (see, for example, Fig.~1 of
Ref.~\citenum{Fairhurst2}), and that may affect how telescopes will be
able to cover them.  Furthermore, coherent reconstruction of arbitrary
burst signals often produces multiple disconnected regions (e.g.,
Fig.~3 of Ref.~\citenum{EMf}) as features in the data stream line up
differently.  The end result in either case is a ``sky map''
representing the estimated probability of the source being at each
position.

Because sky maps may span large areas that are impractical to cover
with most telescopes, the 2009--10 follow-up campaign made the
additional assumption that sources were expected to be in nearby
galaxies or in Milky Way globular clusters.  That is, a catalog of
nearby galaxies (and globular clusters) was used to weight the sky
maps and guess the most likely host galaxies based on distance and
luminosity; those were then imaged.  Monte Carlo simulations indicated
that this approach would significantly improve the chance of imaging
the correct source location~\cite{EMf}, though that conclusion was
model-dependent.  It is not clear whether a galaxy-targeting strategy
will still be useful in the future when the GW detector network will
be reach out to greater distances, encompassing a much larger number
of galaxies per unit solid angle, and where current galaxy catalogs
are much less complete.  It is also not clear that we should expect
all sources to be found in galaxies \cite{Berger2010}.

\subsection{Communication with follow-up observers}  \label{subsec:comm}

After selecting (and validating) an event candidate, the essential
information about the candidate includes the arrival time of the GW
signal at the Earth, a measure of its significance or FAR, and a sky
map describing its apparent location.  The type of signal (e.g.,
binary merger vs.\ burst) and some parameters such as component masses
or burst frequency and duration may also influence the observing
strategy, if they can be determined reasonably accurately from the GW
data.  For the 2009--10 run, the LSC and Virgo used the sky map and
galaxy catalog to select (RA,Dec) coordinates for the telescopes to
point at, but in the future we expect to provide the full sky map to
observers.

The 2009--10 follow-up observing campaign was organized in a
relatively short period of time and involved ten partner groups, each
using one or more telescopes.  Communication protocols were adapted to
each partner's existing telescope scheduling system, with the result
that eight different protocols or interfaces were used---although most
were based on either the GCN \cite{GCN2008} notice binary packet
format or the VOEvent \cite{VOEvent} format.  Future operations will
benefit greatly from distributing alerts through one or two standard
infrastructures that are widely used by the transient astronomy
community at that time, such as GCN/TAN or VOEventNet
\cite{VOEventNet}.  One consideration is that after the initial rapid
alert, revised information may become available later---such as a
refined sky map, physical parameters of the apparent source system,
and/or an updated assessment of the significance of the event
candidate---which should be distributed and may influence further
follow-up observations.

\subsection{Impact of follow-up observations}  \label{subsec:results}

The findings of follow-up observations for a given GW candidate,
combined with the original GW data, can hopefully answer the
questions: Is this event real?  Where is it?  What is the source?
Finding a plausible EM transient counterpart should add confidence,
and in some cases may make the difference between a marginal candidate
and a firmly established astronomical event.  However, given the large
sky areas associated with the GW candidate, it is important to
recognize that false associations with unrelated EM transients are
quite possible, even likely.  It will be essential to classify
transients (e.g., by following light curves or with spectroscopy) to
minimize false associations with unrelated events, and to
understand the effectiveness of that classification for a broad range
of transients, in order to get some handle on the ``false association
probability''.  That will directly affect how much additional
confidence the counterpart provides.

Finally, rapid feedback from observers when a firm or possible EM
counterpart is found will be valuable for the GW analysis of the event
as well as for obtaining further EM observations.  It will likely lead
to deeper investigation of GW data quality and a re-assessment of the
confidence in the event.  And, by pinning down a more precise sky
location and/or estimated distance, it can enable improved estimates
of other parameters of the system from the GW data, such as the masses
and orbital inclination of the compact binary progenitor system, which
would sharpen the emerging picture of the event.

\section{SUMMARY: PROSPECTS FOR MULTI-MESSENGER SCIENCE}
\label{sec:future}  

An advanced gravitational wave detector network is currently under
construction and should be operational around the middle of this
decade.  We are preparing now to be able to rapidly analyze the data
and issue alerts for significant GW event candidates within a few tens
of minutes or less, using lessons learned from our first attempt in
2009--10.  Once the first two LIGO detectors and the Virgo detector
are operating with comparable sensitivity, we should be able to
provide sky maps for some candidates with typical areas of tens of
square degrees, although that area depends strongly on the signal
type, strength, and sky location.  Full LIGO/Virgo detector
sensitivity and additional detectors will follow later on and will
greatly improve the rate of detected events and their localization.

The LSC and Virgo are committed to providing rapid alerts to
astronomers interested in following up GW event candidates.  A policy
document \cite{LVReleasePolicy} posted this year outlines the general
philosophy; the details of how best to implement the plan are
currently being worked out.  As described in the policy, during the
initial science run(s) (while the detectors are still being
commissioned), candidates will be shared with partners through
memoranda of understanding intended to ensure coordination and
confidentiality of information during the time when any public
announcements about the first direct detection(s) of gravitational
waves will need to be carefully vetted.  After 4 GW events have been
published, we will begin releasing highly significant triggers
promptly to the public, while partners will continue to have access to
larger data sets for more systematic follow-up campaigns.

The next several years will be an exciting time for doing science with
gravitational waves {\em and} well-established EM observing.  The
relatively large sky areas involved call for strategic planning and
appropriate observing resources in order to find the counterparts and
rule out false associations, but there is a huge potential payoff in
connecting the complementary signals, which should teach us much about
energetic astrophysical events.

\acknowledgments     
 
The corresponding author is grateful for the support of the U.S.\ National Science Foundation through grant PHY-1068549.
The authors gratefully acknowledge the support of the United States
National Science Foundation for the construction and operation of the
LIGO Laboratory, the Science and Technology Facilities Council of the
United Kingdom, the Max-Planck-Society, and the State of
Niedersachsen/Germany for support of the construction and operation of
the GEO600 detector, and the Italian Istituto Nazionale di Fisica
Nucleare and the French Centre National de la Recherche Scientifique
for the construction and operation of the Virgo detector. The authors
also gratefully acknowledge the support of the research by these
agencies and by the Australian Research Council, 
the International Science Linkages program of the Commonwealth of Australia,
the Council of Scientific and Industrial Research of India, 
the Istituto Nazionale di Fisica Nucleare of Italy, 
the Spanish Ministerio de Econom\'ia y Competitividad,
the Conselleria d'Economia Hisenda i Innovaci\'o of the
Govern de les Illes Balears, the Foundation for Fundamental Research
on Matter supported by the Netherlands Organisation for Scientific Research, 
the Polish Ministry of Science and Higher Education, the FOCUS
Programme of Foundation for Polish Science,
the Royal Society, the Scottish Funding Council, the
Scottish Universities Physics Alliance, The National Aeronautics and
Space Administration, the Carnegie Trust, the Leverhulme Trust, the
David and Lucile Packard Foundation, the Research Corporation, and
the Alfred P. Sloan Foundation.

This paper has been assigned the identifier LIGO-P1200070-v2.


\bibliography{bib_8448_25}   
\bibliographystyle{spiebib}   

\end{document}